# Profiling Frailty: A parsimonious Frailty Index from health administrative data based on POSET theory

*Margherita Silan[1], Maurizio Nicolaio[2], Giovanna Boccuzzo[3]*

**ABSTRACT**

Frailty assessment is crucial for stratifying populations and addressing healthcare challenges associated with ageing. This study proposes a Frailty Index based on administrative health data, with the aim of facilitating informed decision-making and resource allocation in population health management. The aim of this work is to develop a Frailty Index that 1) accurately predicts multiple adverse health outcomes, 2) comprises a parsimonious set of variables, 3) aggregates variables without predefined weights, 4) regenerates when applied to different populations, and 5) relies solely on routinely collected administrative data.

Using administrative data from a local health authority in Italy, we identified two cohorts of individuals aged ≥65 years. A set of six adverse outcomes (death, emergency room access with highest priority, hospitalisation, disability onset, dementia onset, and femur fracture) was selected to define frailty. Variable selection was performed using logistic regression modelling and a forward approach based on partially ordered set (POSET) theory. The final Frailty Index comprised eight variables: age, disability, total number of hospitalisations, mental disorders, neurological diseases, heart failure, kidney failure, and cancer. The Frailty Index performs well or very well for all adverse outcomes (AUC range: 0.664-0.854) except hospitalisation (AUC: 0.664). The index also captured associations between frailty and chronic diseases, comorbidities, and socioeconomic deprivation.

This study presents a validated, parsimonious Frailty Index based on routinely collected administrative data. The proposed approach offers a comprehensive toolkit for stratifying populations by frailty level, facilitating targeted interventions and resource allocation in population health management.

**FINDINGS OF THE ARTICLE:**

- Proposal of a validated Frailty Index based only on routinely collected administrative health data.

[1] Margherita Silan, PhD, Department of Statistical Sciences, University of Padua, Padua, Italy, 35121, email: margherita.silan@unipd.it, ORCID *0000-0001-5541-0603*,
*https://sites.google.com/view/margheritasilan/home?authuser=1*

[2] Maurizio Nicolaio, MD, Department of Statistical Sciences, University of Padua, Padua, Italy, 35121, email: maurizio.nicolaio@unipd.it

[3] Giovanna Boccuzzo, PhD, Department of Statistical Sciences, University of Padua, Padua, Italy, 35121, email: giovanna.boccuzzo@unipd.it, ORCID *0000-0003-2143-7730, https://homes.stat.unipd.it/giovannaboccuzzo/en/home-2/*

- The Frailty Index comprises only eight variables, but demonstrates good performance in predicting multiple adverse outcomes.
- Higher levels of frailty are strongly associated with increased risks of death, disability, dementia, and other adverse outcomes.
- The index captures associations between frailty and chronic diseases, comorbidities, and socioeconomic deprivation.
- The use of Partially Ordered Sets (POSET) theory allows us to aggregate dichotomous and ordinal variables without predefined weights.

**INTRODUCTION**

Demographic ageing is one of the most significant phenomena of our time and is destined to profoundly transform the social, economic, and health dynamics of contemporary societies. This phenomenon is gaining increasing importance as a growing number of older adults live an increasing number of years characterised by disability, chronic diseases, or reduced autonomy.

In this context, the health frailty of people in later life emerges as a central element in understanding and addressing the challenges associated with ageing, both from a scientific and practical perspective.

Frailty can be defined as a multidimensional condition that reflects the decrease in an individual's physiological, cognitive, and social reserves, determining a state of vulnerability that amplifies the risk of functional decline and adverse health outcomes (such as falls, hospitalisations, disability, loss of autonomy) [1, 2, 3].

Identification of frailty poses a significant scientific challenge, as it represents a latent construct, that is, a characteristic not directly observable [4]. This difficulty is mainly due to the multiplicity of theoretical perspectives through which frailty has been conceptualised and the consequent heterogeneity of approaches adopted to measure it.

The definitions of frailty are located along a theoretical spectrum that includes, on the one hand, biomedical and physiological models that focus exclusively on the physical dimension of frailty, and, on the other, multidimensional approaches that also consider cognitive, social, and psychological aspects [5, 6]. The multiplicity of theoretical perspectives translates into a plurality of measurement tools, each of which reflects not only the reference conceptual model but also the practical objectives for which the measurement is carried out and the context in which frailty is detected.

The three most recognised and widely adopted paradigms for defining and measuring frailty will be presented.

According to the biomedical paradigm [1], frailty is defined as a biological syndrome of decreased reserve and resistance to stressors, resulting from declines in multiple physiological systems, causing vulnerability to adverse outcomes. This definition highlights the systemic nature of frailty and its impact on the body's ability to respond to stressors [1]. The most widely recognised and used tool for frailty measurement based on the biomedical conceptualisation is the Fried Frailty Phenotype, based on the assessment of five elements: unintentional weight loss, exhaustion, weakness, slowness, and low physical activity.

In the cumulative deficit paradigm [7], frailty is defined as *a state of chaotic disorganisation of physiological systems that can be estimated by assessing functional status, diseases, physical and* *cognitive deficits, psychosocial risk factors, and geriatric syndromes with the aim of building as complete a picture as possible of risk situations for adverse events* [8]. Several measures have been proposed in the literature, based on the accumulation of deficits perspective [9-14]. The Frailty Index (FI) is one of the most widely used: it uses a comprehensive list of potential deficits [9], which can number between 30 and 70 or more, depending on the availability of the data and on the study design.

Finally, the bio-psycho-social paradigm recognises that frailty is a multidimensional syndrome influenced by a complex interplay of various determinants. Gobbens and coauthors define frailty as a dynamic state that affects an individual who experiences losses in one or more domains of human functioning (physical, psychological, and social), which are caused by the influence of a variety of variables and increase the risk of adverse outcomes [2]. Presented by Gobbens and coauthors in 2010, the Tilburg Frailty Indicator (TFI) is based on 15 items collected in three domains using a self-administered questionnaire [15]. These domains include physical components, psychological factors, and social elements.

The frailty indices proposed in the literature start from different information needs: some focus on specific groups of people, such as the institutionalised older people, or those who have undergone specific interventions, or with certain chronic conditions. In this case, one relies on ad-hoc surveys, in which a lot of specific data is collected. In this instance, it is clearly possible to adopt any theoretical approach to measure frailty, as the data are collected specifically for this purpose.

Other studies, and this is our case, are instead aimed at the entire population and are necessarily based on census or administrative data. In this context, the main objective is to stratify the population according to certain health characteristics or care needs.

The importance of stratifying populations based on frailty levels is underscored by various regulations and the potential for targeted interventions. Frailty is associated with higher healthcare costs, greater risk of adverse events during and after surgery, markedly poorer quality of life, and increased burden for family caregivers [16]. In Italy, the National Health System has implemented measures to address this problem, such as the 2016 National Chronicity Plan, which aims to stratify the population based on clinical and socio-family factors. In 2022, Decree No. 77 introduced additional innovations within the National Recovery and Resilience Plan, focussing on measuring frailty and risk-based stratification [17]. This reform established new models for territorial healthcare, recommending the implementation of population measurement and stratification systems based on risk. The goal is to create a platform that contains information on the characteristics of the assisted population, the prevalence of chronic diseases, and the frail population. The decree emphasises that population stratification methodologies are fundamental tools for defining effective proactive healthcare strategies, optimising multidisciplinary treatment, personalising care and reducing inappropriate interventions and associated costs.

In recent years, Frailty Indexes based on administrative data have become fundamental role in assessing the risk of adverse health outcomes in older adults and the frail population.

These indexes are based on the validity of the criterion [18]: the operational definition must correctly classify individuals with respect to adverse health outcomes, such as mortality, institutionalisation, long hospital stays, and falls.

In the literature, there are several proposals for *Electronic Frailty Indices* based on administrative data, which are based on Electronic Health Records. They can be broadly divided into two approaches. Most studies are based on a modelling approach, in which logistic regression models (or more advanced statistical techniques, such as LASSO regularisation) allow the identification of the most relevant variables to predict adverse outcomes. The results of the model enable the calculation of the probability that a given individual may experience an adverse event. These probabilities are used as a true FI [19-27].

A second approach is to construct the FI using the *accumulation of deficits* proposed by Mitnitski et al [28-30].

Once frailty scores are assigned to subjects, the most common approach to assess the validity of the proposed measure involves evaluating the goodness of the index through the analysis of its predictive capacity with respect to different health outcomes. The measure used is the Area Under the ROC (Receiver Operating Characteristic) curve, known as AUC (Area Under the Curve).

Our work aligns with the approach that aims to predict adverse outcomes while addressing the limitations of both methods. The *modelling* approach presents two main limitations:

1. It typically predicts one or at most two adverse outcomes, considering the dependent variable to have experienced at least one of the two outcomes (typically mortality and hospitalisation, or mortality and disability). If it considers more outcomes, it models the one that occurred first. As frailty can lead to various adverse health outcomes, it would be desirable to predict multiple outcomes simultaneously.
2. It provides indexes based on the linear combination of variables weighted by regression coefficients calculated on specific populations. The underlying assumption is that the weights estimated for one population can be generalised to other contexts.

The deficit accumulation approach essentially constructs a summative indicator, sometimes adding weights to the index components. Criteria can vary. Again, the assumption is that the weights remain constant or need to be recalculated in different contexts.

Finally, in both approaches, many indices are based on large numbers of variables from both administrative and survey data sources, making their actual calculation in settings such as healthcare providers very burdensome.

In this work, we revisit and expand on the methodological approach proposed by Silan et al. [31, 32], which offers a more robust and comprehensive framework for assessing frailty assessment using administrative data.

The overarching goal of this work is to offer a comprehensive toolkit for researchers and healthcare professionals seeking to stratify populations by frailty level using administrative health data. By providing a validated and methodologically sound approach to frailty assessment, our objective is to contribute significantly to the field of population health and facilitate more informed decision making in the allocation of healthcare resources and intervention planning.

In detail, our proposal aims to propose a FI:

1. That is a good marker of frailty and allows the prediction of multiple negative outcomes in the entire population.
2. Composed of a parsimonious set of variables, so that its construction is not too laborious for public health operators.
3. In which the variables are aggregated using a method that does not need the inclusion of weights.
4. That regenerates when applied to different populations.
5. That is based on administrative data available in each Local Health Unit (thereby avoiding information not present in administrative registries).

The points listed represent, to our knowledge, aspects of innovation with respect to the existing literature and resolve both conceptual limitations (providing only one outcome, using fixed weights) and operational limitations (simplicity of calculation, use of data available in each local health authority, regeneration of the index when applied to different populations).

From a methodological standpoint, we base our approach on the theory of partially ordered sets (POSET). We provide a thorough validation process, in the construction steps, in the application, and in the performance assessment of the Frailty Index (FI). The POSET approach is particularly effective for composing measures based on dichotomous and ordinal variables, and does not require the use of weights.

The article is organised as follows: the *Methods* section describes the path to produce the frailty index, focussing on the strengths and limitations of the POSET approach and on the robustness analysis. The *Results* session describes and validates the proposed FI, showing its performance in terms of predicting negative health outcomes, stratifying the population and underlining its relationships with chronic diseases, comorbidities, disability and socioeconomic deprivation. The last two sections, *Discussion* and *Conclusions and future work*, provide several considerations about key findings, strengths, and limitations of our proposal, the implications in the use of health administrative data, policy considerations, and future work.

## METHODS

### *Context*

The National Health Service (NHS) in Italy is a regionally-based healthcare system. It provides universal coverage free of charge at the point of service. The central government ensures compliance with the general objectives and fundamental principles of the NHS, defines a national package of statutory benefits set of *essential levels of care* for the population, and allocates national funds to regional authorities. Regional governments (19 regions and two autonomous provinces) are responsible for organising and delivering healthcare services through a network of local health authorities. The latter have catchment areas that are often defined by the boundaries of provinces (institutional entities that group together several municipalities in a given region). Depending on the region, local health authorities to public hospitals and accredited private clinics.

### *Data sources and variables*

The analysed populations include subjects residing in the province of Padua (Veneto Region, Italy), assisted by the Local Health Authority "ULSS6 Euganea". From the health registry, we defined two cohorts that included subjects 65 years or older who received assistance during the periods 2016-2018 and 2017-2019. We identified 213,689 and 216,757 subjects, respectively (of which 205,004 are common to both cohorts), aged at least 65 years as of January 1, 2018 and 2019. The variables used to construct the FI were observed in the first two years of each period. The outcomes were observed in the third year.
We used eight administrative health databases:

1. Regional health registry, which also includes the death registry, necessary to identify the reference population;
2. Hospital discharge records, containing all hospitalisations and related information, such as duration, primary diagnosis, and up to five concomitant diagnoses;
3. Emergency room admissions (ER), containing all accesses to the ER and related information such as priority level, primary diagnosis, and up to 4 concomitant diagnoses;

4. Territorial psychiatry, which includes the services provided, the primary diagnosis, and up to two concomitant diagnoses;
5. Integrated home care, which contains all beneficiaries of this service, the number and duration of services;
6. Ticket exemptions, which includes beneficiaries of exemptions with information on pathologies or economic conditions;
7. Territorial pharmaceuticals, including the list of prescribed drugs;
8. Outpatient services refarding older people dependent on services, including the list of residential care services, home care, and other forms of long-term support provided.

The information from all sources under analysis was coded, combined and released to us after an anonymisation process, with the aim of making personal data unidentifiable. Through administrative flows, it was possible to retrieve information related to nine outcomes and 75 frailty markers.

## Process for constructing the Frailty Index

The construction of a composite index is a complex process that requires a number of decisions and assessments, both methodological and otherwise, depending on the context, objective and available data. Figure 1 summarises all the steps needed to construct the FI.

Figure 1: Steps to construct the Frailty Index.

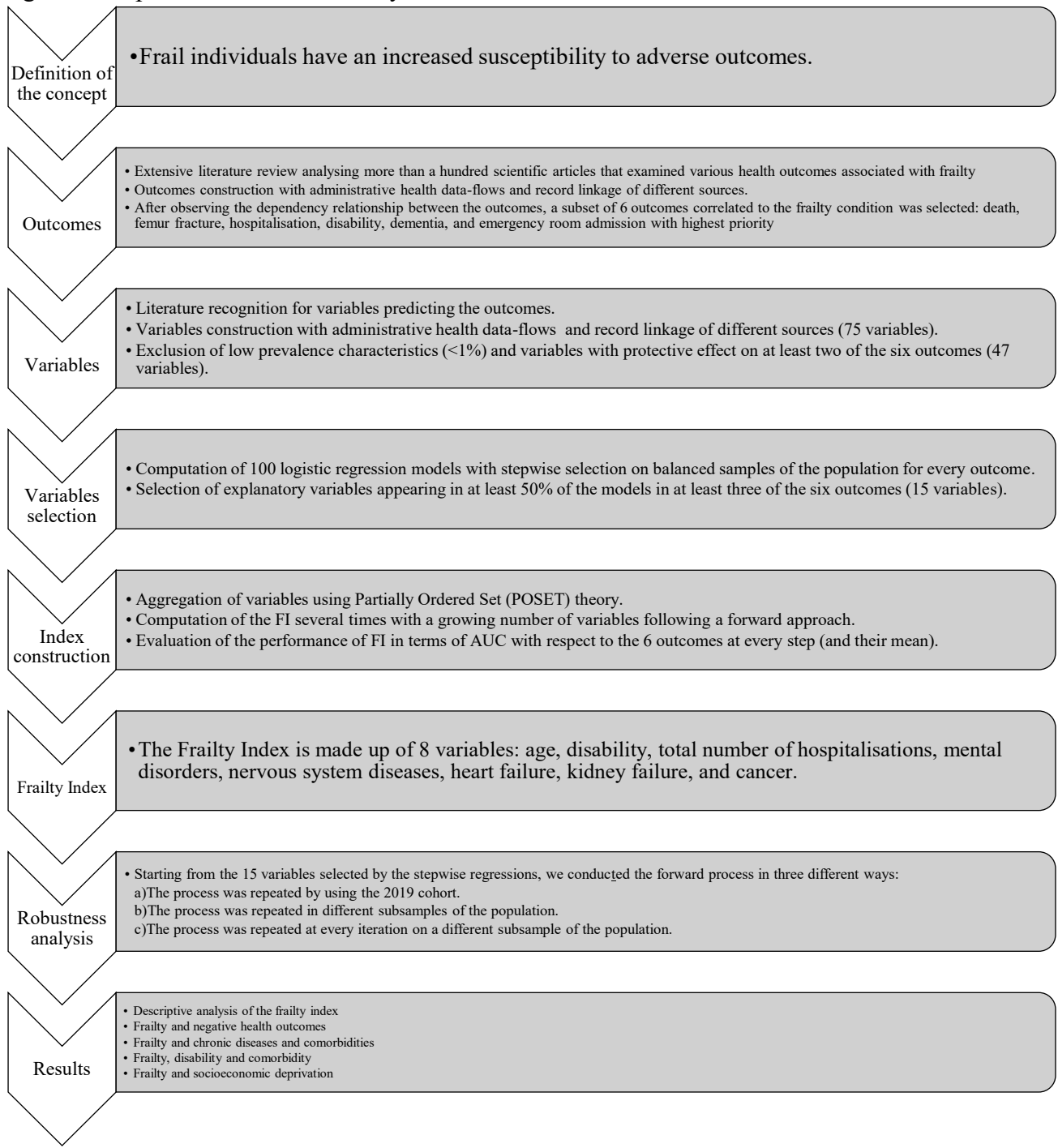

## Choice of adverse outcomes

A recognised approach to defining and measuring frailty is through the prediction of negative health outcomes [18]. The choice of the set of outcomes to consider is a crucial first step in this process.

Although research focusses on a single outcome at a time (or on the presence of at least one of several outcomes considered), the multidimensional nature of frailty requires considering multiple outcomes simultaneously. However, including too many outcomes can unnecessarily complicate the analysis. To address this, an extensive literature review was conducted, analysing more than a hundred scientific articles that examined various health outcomes associated with frailty [33]. Through factor analysis and graphical modelling techniques, the relationships and interconnections between these outcomes were thoroughly investigated. This systematic approach allowed for the identification of a fundamental and parsimonious set of six representative outcomes: death, femur fracture, hospitalisation, disability, dementia, and emergency room admission with highest priority (equivalent to the Italian *red code* triage system for life-threatening conditions). For the outcomes of dementia and disability, only the onset of these two conditions is considered an event. Subjects who have already been diagnosed with dementia and disability in the first two years cannot therefore present these two (new) characteristics in the year of outcome prediction. This concise yet comprehensive set captures the critical events and chronic conditions that characterise frailty. The outcomes are observable and accessible to all subjects through health administrative flows.

*Identification of the variables that compose the Frailty Index*
Variables that identify frailty are those factors that influence an individual's health status and lead to adverse outcomes. We identified 75 variables as potential markers of frailty. This list was obtained based on the markers most used in the literature to build FI and is composed of all those that influence even a single outcome at a time. Some of these variables also assume a dual role: In the literature, some phenomena are considered both among the determinants of frailty and among the outcomes. This occurs, for example, for disability, dementia, fractures, and hospitalisations. Continuous and count variables were grouped into meaningful categories using repeated classification trees in subsamples to correctly classify individuals who experienced the 6 outcomes.
The selection was carried out according to the following procedure:
1. Calculation of the prevalence of each condition. Exclusion of variables with a prevalence in the population lower than 1%;
2. Calculation of the odds ratios between each variable and each adverse outcome. Removal of variables with protective effects on at least two of the six considered outcomes;
3. Estimation of 100 logistic regression models via stepwise selection approach on balanced subsamples of the whole population, for each outcome. Selection of explanatory variables that appear in at least 50% of the models in at least three of the six outcomes considered.

At the end of this selection step, 15 variables are selected and considered in the following step. This group of variables is the most effective in predicting each of the frailty-related outcomes when considered individually.

*Index construction with Partially Ordered Set (POSET) theory*

Variables are aggregated by exploiting Partially Ordered Sets (POSET) theory. The method orders subjects based solely on their observed characteristics, avoiding subjective elements and leveraging the ordinal nature of the variables. The POSET approach is applicable to any type of variable but lends itself particularly well to ordinal and dichotomous variables. The POSET method considers each subject's profile, i.e. the set of values that a subject takes on the variables considered, and generates a measure called Average Rank (AR) through comparison of all profiles. Each subject is associated with a specific profile and assigned a normalised AR value indicating his/her frailty level. The AR arranges individuals according to their relative frailty, making it ideal for identifying the frailest members of a specific population. However, AR is population-dependent, as it depends on the set of profiles observed in the specific population, and cannot be interpreted in absolute terms or used to compare individuals in different populations. The AR value does not determine whether a subject is frail or not but rather allows for ranking a population based on the characteristics of its members according to frailty levels. Additional details about the method can be found in Silan et al. (2019 and 2022) [31, 32].

Starting with the subset of 15 frailty markers, a second step of variable selection is performed based on POSET theory and with a forward approach. Reducing the number of variables allows one to maintain computational efficiency without compromising the accuracy of outcome predictions.

This approach is effective because including a new variable does not always guarantee an improvement in the predictive performance of the index due to increased entropy and the number of incomparable profiles.

To measure the predictive capacity for the six outcomes, the area under the ROC curve (AUC) is used, calculated separately for each adverse outcome. For the *onset of disability* and *onset of dementia outcomes, the* AUCs are calculated in the population of subjects who, in the two years preceding the outcome year, do not have a disability or a dementia diagnosis, respectively. Variable selection is based on the mean of AUCs for the six outcomes, following this process with *forward* logic:

1. Step 1: The two variables are chosen, which, when combined, maximise the prediction performance of the six outcomes.
2. Steps 2 to n-1: At each step, the variable that, when combined with those already selected, maximises the mean of AUCs is added.
3. Step n: The process ends because none of the remaining variables leads to further improvement in predictive performance. The variables selected up to the previous step (n-1) are those that make up the final FI.

Therefore, the final FI is made up of the following 8 variables: age, disability, total number of hospitalisations, mental disorders, nervous system diseases, heart failure, kidney failure, and cancer[4]. With eight variables in a population of 213,689 individuals (878 different observed profiles), the computational time is around 3 minutes (CPU time: 175.23 seconds[5]). The algorithms for constructing these variables are provided in the Supplementary Material, while the list and algorithms for the others are available upon request. To calculate the confidence intervals (CIs) of the areas under the ROC curve (AUC), we used the DeLong method [34]. In R, this was implemented using the *ci.auc* function of the *pROC package.*

All in all, the POSET approach allows us to fully exploit the ordinal information contained in the dataset with the need for a small set of variables. Having these variables selected by considering simultaneously all the frailty related outcomes, the result is that the FI has a good predictive performance with respect to all the six outcomes (with some differences explained in the *Results*). At the same time, outcomes are only considered as guidance to select meaningful and predictive variables, and they are not considered in the computation of the FI, which generates itself from the parsimonious set of variables without the need for coefficients, weights, or additional hypotheses.

Given the importance of selecting a parsimonious set of variables for the calculation of the FI, this step has been enriched with a detailed analysis of robustness.

Starting from the 15 variables selected by the stepwise regressions, we conducted the forward process in three different ways to ensure that the final set of variables is not contingent on the analysed population:

a) The process was repeated using the 2019 cohort.
b) The process was repeated in different subsamples of the population.
c) The process was repeated at each iteration on a different subsample of the population.

---

[4] The *cancer* variable refers to an active condition currently undergoing treatment, and not to a post-treatment follow-up situation. In other words, the index considers the presence of a neoplasm currently in the diagnostic or treatment phase as a frailty factor, rather than a previous history of cancer in the monitoring phase.

[5] CPU: Intel(R) Core(TM) Ultra 7 155U (1.70 GHz); CORE: 12; RAM: 32 GB.

For point b), we adopted the following approach: first, we divided the total population into four subsamples. Then, we conducted four iterations of the process, each time excluding one of the subsamples to create a distinct population for analysis. The entire procedure, including the division into subsamples and the four iterations, was repeated twice to ensure robustness. At point c), in each iteration, 10% of the total population was excluded and subsequently reintegrated, thus altering the population structure at each step.

**RESULTS**

*Robustness analysis*

The results of the robustness analyses performed to verify the reliability of the variable selection are reported in Table 1. This table shows the selected variables and the means of AUCs for both the proposed FI and the various scenarios analysed and compared.

Applying the same methodology to the core set of 15 variables identified for the 2018 cohort in the 2019 cohort (scenario a) yields the exact same set of selected variables, with a higher overall mean of AUCs compared to the FI referred to the 2018 cohort.

Regarding the selection of population subsamples (scenario b):
- In both tests (b1 and b2), three of four samples select the same variables.
- In one of the four folds, the process stops before including cancer among the FI components.

When selecting a different population subsample at each iteration (scenario c), the process includes the same variables, except for cancer and kidney failure.

The variables selected for the FI are observed to remain substantially consistent, with minimal variations, despite the resampling of the analysed population and changes in the observation period.

Table 1: Selected variables and mean of AUCs for the proposed FI and the scenarios considered in the robustness analysis.

| Variable | Frailty Index | (a) | (b1) | | | | (b2) | | | | (c) |
|---|---|---|---|---|---|---|---|---|---|---|---|
| | | | 1st fold | 2nd fold | 3rd fold | 4th fold | 1st fold | 2nd fold | 3rd fold | 4th fold | |
| Age | X | X | X | X | X | X | X | X | X | X | X |
| Disability | X | X | X | X | X | X | X | X | X | X | X |
| Total number of hospitalisations | X | X | X | X | X | X | X | X | X | X | X |
| Mental disorders | X | X | X | X | X | X | X | X | X | X | X |
| Nervous system diseases | X | X | X | X | X | X | X | X | X | X | X |
| Heart failure | X | X | X | X | X | X | X | X | X | X | X |
| Kidney failure | X | X | X | X | X | X | X | X | X | X | |
| Cancer | X | X | X | X | X | | X | | X | X | |
| Mean of AUCs | 0.77 | 0.78 | 0.78 | 0.78 | 0.77 | 0.77 | 0.77 | 0.77 | 0.77 | 0.78 | 0.77 |

*Descriptive analysis of the Frailty Index (FI)*

The FI values, obtained using the POSET method, are transformed into a normalised scale, i.e., reported to a common interval (between 0 and 1). This normalisation process simplifies the interpretation of the frailty level associated with individuals. Figure 2 shows the distributions of FI referring to the 2018 and 2019 cohorts, respectively. Table 2 reports some descriptive statistics of the FI.

We can observe that the distributions of FI are very similar in the 2018 and 2019 cohorts and highly asymmetric, characterised by a clear concentration of individuals with low FI values and a minority of subjects with high FI. Indeed, the median of FI has a value of 0.06 and 25% of subjects with high levels of frailty have a value equal to or greater than 0.18. This distribution reflects the demographic and health reality of the analysed population: most individuals have relatively good or moderately compromised health conditions, while only a small portion of the population shows high levels of frailty. This aspect underscores the importance of interpreting the level of frailty in relation to the overall distribution of FI in the population, rather than in absolute terms. Even apparently moderate values can indicate compromised health conditions compared to most individuals. Subjects with higher FI are characterised by more severe cases, advanced age and the simultaneous presence of multiple chronic diseases.

Table 2: Descriptive statistics of FI attributed to subjects in the 2018 and 2019 cohorts.

| Population | Minimum | 1st Quartile | Median | Average | 3rd Quartile | Maximum |
|---|---|---|---|---|---|---|
| 2018 cohort | 0 | 0.01 | 0.063 | 0.125 | 0.186 | 1 |
| 2019 cohort | 0 | 0.01 | 0.061 | 0.123 | 0.182 | 1 |

Figure 2: Distribution of FI calculated in the 2018 (left) and in the 2019 (right) cohort.

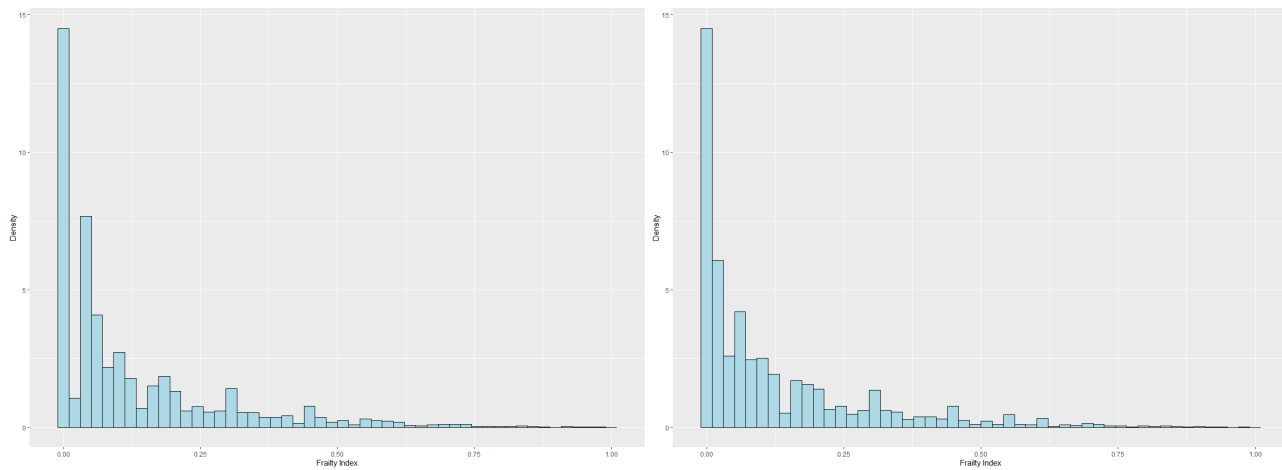

The FIs calculated for the subjects present in both the 2018 and 2019 cohorts show a Spearman correlation of 0.88. This result reflects an overall stability in the level of frailty of subjects over time. However, the analysis of subgroups reveals significant differences:

- Subjects who maintain the same profile (with stable and unchanged characteristics and risk factors over the two years) in both cohorts show a higher correlation, close to 1, indicating strong consistency in their level of frailty, as expected given the absence of significant variations in their conditions.
- On the contrary, subjects who change profile in the 2019 cohort show a correlation of 0.73. This result is consistent with the idea that the change in profile reflects an evolution in the health condition or general situation of the subjects, influencing their FI.

Comparing common profiles in the two cohorts allows us to assess to what extent changes in the structure of the population (around 90% of profiles observed in the 2018 cohort are observed also in the 2019 cohort) lead to variations in the FI that are not caused by a modification in the physical conditions of the considered older adults. From this analysis, we may appreciate that in large populations structural changes between subsequent years may be negligible at individual level, having correlation close to 1 for subjects who maintain the same profile.

Figure 3 shows the means of the FI values and their 95% confidence intervals by age class and sex. The distributions for both genders follow a similar pattern, with FI values increasing with age. In younger age classes, men tend to have slightly higher mean FI values compared to females. As age increases, the gap between the male and female FI values narrows.

Figure 3: Mean and 95% CI for FI by age class and sex.

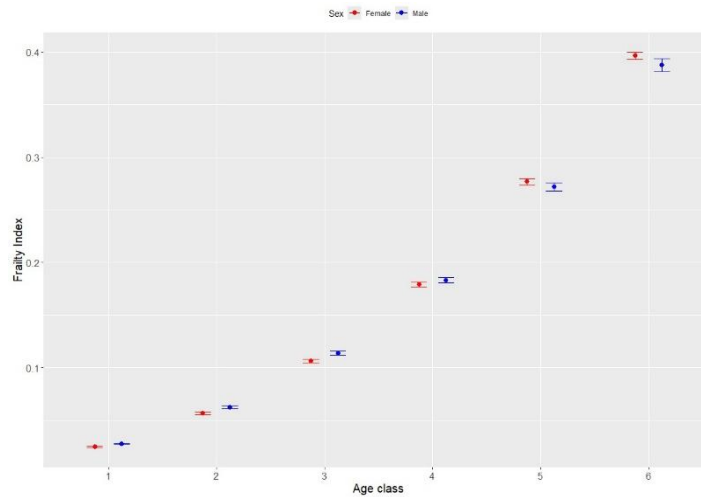

*Population stratification based on the Frailty Index*

As population stratification is a key goal of the FI, we investigate the effectiveness of population stratification with different thresholds. Our analysis focusses on observing the occurrence of frailty-related outcomes across different strata, allowing us to evaluate the effectiveness of the FI as a measure to identify and categorise levels of frailty within the population. As a first tentative, the population was divided into four groups based on the quartiles of FI values. Table 3 shows the FI values corresponding to each quartile and the number of different profiles present. The division of the population into quartiles results in a slightly irregular distribution compared to the expected percentage, having a total of 872 profiles considered and a consequent step-like distribution.

From the table, it can be observed that the subjects in the last two quartiles have a more heterogeneous distribution of profiles, associated with more severe health conditions and less frequent situations. On the contrary, the first quartile includes a significant proportion of subjects having the same profile: in particular, these are individuals aged between 65 and 69 years who do not present any of the diseases or conditions considered for calculating the FI.

Table 3: Number of different profiles and extremes of each quartile of the FI distribution in the 2018 cohort.

| Quartile | FI range | Number of profiles |
|---|---|---|
| 1 | [0 - 0.010] | 2 |
| 2 | (0.010 - 0.063] | 10 |
| 3 | (0.063 - 0.186] | 63 |
| 4 | (0.186 - 1] | 797 |

*Distribution of each adverse outcome by FI levels*

The objective of this analysis is to verify whether high FI values (computed using 2016 and 2017 data) are associated with a higher probability of experiencing the six frailty outcomes considered in the following year (2018).

In previous analyses, we considered the onset of two frailty markers among outcomes: disability and dementia. However, in this analysis, we include the prevalence of these two conditions in the outcome count. This approach is necessary because, being chronic and often irreversible conditions, their presence offers a more accurate representation of the subject's overall frailty status compared to onset alone. Otherwise, we would not be able to adequately represent those who are already disabled and/or those who have a diagnosis of dementia.

The first column of Table 4 shows the percentages of occurrence of each outcome during 2018. The last four columns of Table 4 show the prevalence of outcomes in each quartile. The prevalence of outcomes by quartile allows us to assess to what extent each outcome is observed in different frailty ranges, highlighting how the probability of experiencing such outcomes varies with the level of frailty.

The prevalence of outcomes in quartiles underlines that among subjects in the last quartile, 70.11% had a disability (new or pre-existing) in 2018, 30.02% had at least one hospitalisation during the year, and 11.82% died. High prevalences are also recorded in the third quartile, with 34.96% of subjects with disabilities and 18.10% having undergone at least one hospitalisation. In the first quartile, that is, among subjects considered less frail, all prevalences are below 1%, except hospitalisation, which involves 8.99% of subjects, and disability, present in 1.54% of cases.

Table 4: Prevalence of outcomes in the whole 2018 cohort and in the FI quartiles.

| *Outcomes* | Prevalence in 2018 cohort | 1st Quartile | 2nd Quartile | 3rd Quartile | 4th quartile |
|---|---|---|---|---|---|
| Death | 3.63% | 0.34% | 0.77% | 2.00% | 11.82% |
| Access to the emergency room with highest priority | 1.29% | 0.21% | 0.43% | 0.87% | 3.80% |
| Hospitalisation | 17.11% | 8.99% | 13.08% | 18.10% | 30.02% |
| Disability (prevalence) | 27.10% | 1.54% | 8.05% | 34.96% | 70.11% |
| Dementia (prevalence) | 3.37% | 0.10% | 0.47% | 1.86% | 11.46% |
| Femur fracture | 0.58% | 0.09% | 0.23% | 0.55% | 1.54% |

Table 5 reports the distributions of each outcome observed in different quartiles of FI for the 2018 cohort. The distributions of the outcomes in the quartiles allow us to verify whether a higher level of frailty translates into a higher probability of experiencing adverse outcomes.

We can observe that, in general, all outcomes occur more frequently in the upper quartiles of FI. The only outcome that shows a less marked difference is hospitalisation, with approximately 1 in 3 hospitalised subjects falling in the first two quartiles. Death, dementia, and disability are the outcomes that show a more concentrated distribution in the upper quartiles, with more than 90% of subjects who experienced one of these events falling in the last two quartiles. In particular, more than 80% of deceased subjects and 84.08% of subjects with dementia are in the group of subjects with the highest FI.

Table 5: Distribution of outcomes in the FI quartiles of the 2018 cohort.

| Outcomes | 1st e 2nd Quartile | 3rd Quartile | 4th Quartile | Total |
|---|---|---|---|---|
| Death | 7.98% | 11.53% | 80.49% | 100% |
| Access to the emergency room with highest priority | 13.09% | 14.14% | 72.77% | 100% |
| Hospitalisation | 34.46% | 22.16% | 43.38% | 100% |
| Disability (prevalence) | 9.02% | 27.03% | 63.95% | 100% |
| Dementia (prevalence) | 4.34% | 11.58% | 84.08% | 100% |
| Femur fracture | 14.65% | 19.74% | 65.61% | 100% |

*Distribution of the number of adverse outcomes by FI levels*

In addition to verifying how individual outcomes are distributed across the FI quartiles of the population, it is also of interest to analyse how subjects are distributed based on the number of outcomes experienced during the year 2018. The underlying hypothesis is that the FI is higher for those who experience many outcomes. In this case, as well, we will consider the prevalence of disability and dementia in the outcome count, not just the onset cases.

Table 6 shows the distributions of the number of outcomes observed in 2018 in different quartiles of FI in the population. Most subjects, equal to 63.13% of the population, did not experience any of the considered outcomes, reflecting a relatively good health status. Only 1.09% of the population experienced four or more outcomes, representing the cases of the greatest complexity and frailty. We can observe that subjects with a high number of outcomes are allocated in most cases to the last quartile, and the remaining subjects are mostly identified in the third quartile. On the other hand, individuals who did not experience any event in 2018 are predominantly distributed in the first two quartiles (about 3 out of 4 subjects); only 8.53% of these subjects have a FI that places them among those with a high level of frailty.

Table 6: Distribution of the number of outcomes in the FI quartiles of the 2018 cohort.

| Number of outcomes | In 2018 cohort | 1st e 2nd Quartile | 3rd Quartile | 4th Quartile | Total |
|---|---|---|---|---|---|
| 6 | 0.01% | 0% | 11.11% | 88.89% | 100% |
| 5 | 0.17% | 0.81% | 5.05% | 94.14% | 100% |
| 4 | 0.91% | 2.84% | 8.48% | 88.68% | 100% |
| 3 | 2.75% | 6.05% | 12.20% | 81.75% | 100% |
| 2 | 7.27% | 10.73% | 21.85% | 67.42% | 100% |
| 1 | 25.76% | 26.74% | 29.83% | 43.43% | 100% |
| 0 | 63.13% | 73.64% | 17.83% | 8.53% | 100% |

*Distribution of adverse outcomes in the most frail subjects*

The previous results clearly show that subjects belonging to the highest quartiles of FI have a higher probability of experiencing the outcomes considered, indicating a higher level of frailty. However, in the case of targeted interventions, local health authorities may be forced to concentrate resources on a narrower part of the population, due to the costs associated with such initiatives. This could involve focussing on a limited percentage of subjects, for example 1% or 10% of the assisted population, with the need to identify cases with the highest levels of frailty. Therefore, it is interesting to examine more closely the subjects with the highest FI.

Table 7 shows the prevalence of adverse outcomes experienced in 2018 in relation to different thresholds of the last percentiles of FI in the population. Among subjects who, according to the FI based on information recovered in 2016 and 2017, belong to the 1% most frail, a marked concentration of adverse outcomes is observed in 2018. In this group, 97.36% have a disability (new or pre-existing), 51.65% have undergone at least one hospitalisation, and 34.77% have died. The prevalence of dementia is also significant, with about one-third of subjects affected by this condition. Among the outcomes, femur fracture is the least frequent, having affected only 1.67% of the most frail subjects.

Considering larger samples, such as the 5% or 10% of the most frail subjects, a progressive decrease in the percentage frequency of all outcomes is noted. This phenomenon is explained by the inclusion of subjects with slightly lower levels of frailty, who tend to experience fewer outcomes compared to the 1% most frail. For example, the prevalence of deaths in 2018 drops from 34.77% to 18.21% when considering the 10% most frail in 2016-2017. However, the prevalence of disability remains very high, affecting 89.18% of subjects in this range. An exception to this trend is the femur fracture, whose frequency remains stable, indicating a less concentrated distribution among subjects with the highest levels of frailty.

Table 7: Prevalence of outcomes in relation to different thresholds of the last percentiles of FI in the 2016-2018 cohort.

| *Outcomes* | Thresholds | | | | |
| --- | --- | --- | --- | --- | --- |
| | **25%** | **10%** | **5%** | **2%** | **1%** |
| Death | 11.82% | 18.21% | 23.31% | 30.35% | 34.77% |
| Access to the emergency room with highest priority | 3.80% | 5.68% | 7.04% | 9.16% | 10.11% |
| Hospitalisation | 30.02% | 36.99% | 42.35% | 48.32% | 51.65% |
| Disability (prevalence) | 70.11% | 89.18% | 91.53% | 96.45% | 97.36% |
| Dementia (prevalence) | 11.46% | 19.9% | 26.55% | 32.62% | 33.7% |
| Femur fracture | 1.54% | 1.73% | 1.67% | 1.73% | 1.67% |

The entire set of analyses, which focused on observing the occurrence of frailty-related outcomes in various strata, was subsequently repeated for the 2019 cohort. The results obtained from this replication were consistent with the findings observed in the 2018 cohort, reinforcing the robustness of our observations over different time periods.

*Reproducibility: Frailty and adverse health outcomes over time and in different populations*

The validity of the FI is measured in terms of its predictive abilities with respect to the six outcomes.

In 2018, we observed a high predictive performance for death (with an AUC of 0.854), and a good performance also for access to the emergency room with the highest priority, femur fracture, and the incidence of dementia and disability. The performance for the remaining outcome, hospitalisation, is slightly worse; this is probably because it is a less specific outcome. Indeed, it is more common to be hospitalised even for individuals who are not specifically frail. Although it is a more general outcome, it is still important to consider it because it represents an event related to a more vulnerable and precarious health condition, especially among older people.

Although we expect good predictive performance for the 2018 cohort by construction (FI calculated with 2016-2017 data and outcomes observed in 2018), it is relevant to measure its performance in the 2019 cohort (FI computed with 2017-2018 data and outcomes observed in 2019). The only necessary assumption for the computation of this FI in the 2019 cohort concerns the choice of the eight relevant variables. The variables were calculated from data collected in 2017 and 2018, while the outcomes were observed in 2019. Table 8 shows the comparisons of the AUCs of the outcomes calculated in the 2018 and 2019 cohorts.

The only significantly different AUCs are related to the outcome of disability onset. This result led to an overall increase of AUCs in the 2019 cohort, with high performance even in a population different from that used to construct the FI.

Table 8: AUC and 95% CI of the six outcomes for the 2018 and 2019 cohorts.

| *Outcome* | AUC 2018 cohort | AUC 2019 cohort |
|---|---|---|
| Death | 0.854 (0.850 – 0.858) | 0.854 (0.850 – 0.858) |
| Access to the emergency room with highest priority | 0.805 (0.797 – 0.814) | 0.812 (0.804 – 0.821) |
| Hospitalisation | 0.664 (0.661 – 0.667) | 0.664 (0.661 – 0.667) |
| Disability (incidence) | 0.749 (0.743 – 0.755) | 0.792 (0.787 – 0.797) |
| Dementia (incidence) | 0.805 (0.797 – 0.812) | 0.806 (0.799 – 0.814) |
| Femur fracture | 0.765 (0.754 – 0.777) | 0.758 (0.747 – 0.770) |

Table 9 shows the AUCs and 95% confidence intervals for the FI calculated separately in subgroups of men and women in the 2018 cohort; it also reports the overall predictive performance for the entire population, obtained considering the FI calculated in distinct subgroups.

No significant differences in predictive abilities are observed between the FI calculated separately for men and women and the one calculated for the entire population. This result is confirmed by the extremely high correlation between the FI calculated in the entire population and those determined separately for the two genders, which is equal to 0.998. This correlation suggests that, despite differences in profile composition between gender subgroups, FI values remain substantially stable and consistent.

Table 9: AUC and 95% CI calculated in the entire population and separately by gender using FIs calculated separately for men and women.

| *Outcome* | AUC Entire population | AUC Male subpopulation | AUC Female subpopulation |
|---|---|---|---|
| Death | 0.855 (0.851 – 0.859) | 0.846 (0.839 – 0.853) | 0.862 (0.857 – 0.867) |
| Access to the emergency room with highest priority | 0.806 (0.798 – 0.814) | 0.783 (0.770 – 0.796) | 0.825 (0.816 – 0.835) |
| Hospitalisation | 0.665 (0.662 – 0.668) | 0.669 (0.665 – 0.674) | 0.665 (0.660 – 0.669) |
| Disability (incidence) | 0.749 (0.743 – 0.754) | 0.742 (0.732 – 0.751) | 0.753 (0.746 – 0.760) |
| Dementia (incidence) | 0.805 (0.797 – 0.812) | 0.804 (0.791 – 0.816) | 0.802 (0.793 – 0.811) |
| Femur fracture | 0.763 (0.752 – 0.775) | 0.791 (0.768 – 0.815) | 0.746 (0.733 – 0.759) |

Thanks to collaboration with the Piedmont Regional Epidemiology Service, the indicator was also calculated for the population of assisted individuals recorded in the Piedmont Region Health Information System. The results obtained were consistent with those observed in the two cohorts in the province of Padua. Figure 4 includes spider charts that illustrate the indicator performance, in terms of AUCs.

Figure 4: Spider graphs with indicator's performance in the Piedmont Region (2018).

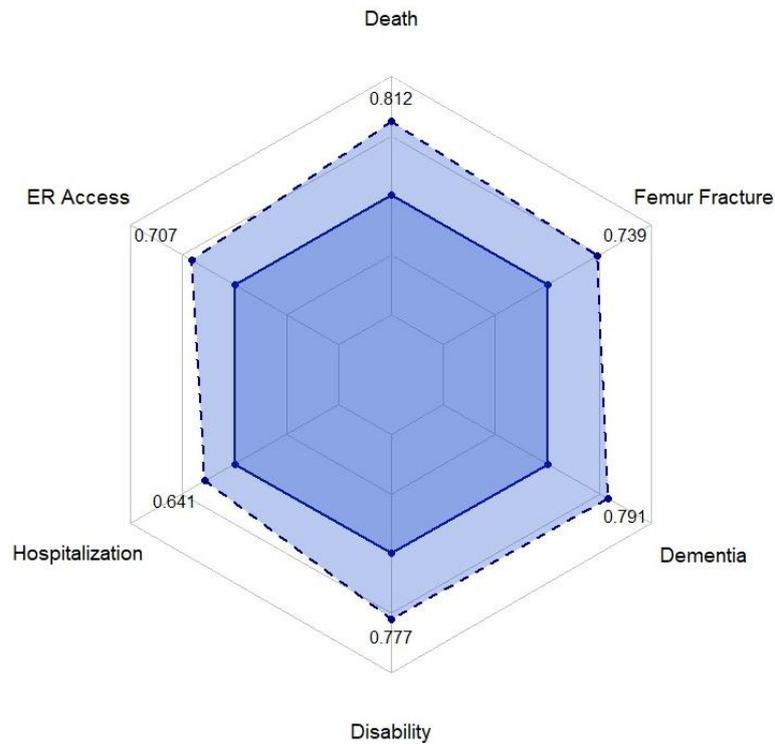

*Frailty and chronic diseases*

To better understanding of the relationship between chronic diseases and the level of frailty in older adults, we observe the values of FI in presence of chronic diseases, providing useful information to manage and prevent vulnerable conditions in the population.

Figure 5 shows, for each disease, the distribution of FI values separately for subjects affected by the condition and those unaffected. Specifically, it represents the median, first quartile, and third quartile, allowing for a direct comparison of the FI distribution between the two groups. It can be observed that for most diseases and conditions, the median FI is consistently higher in subjects with the disease compared to those without it, even if these illnesses did not enter directly into the calculation of the indicator.

Indeed, FI shows a strong relationship not only with diseases that directly compose it, such as renal failure and heart failure, or indirectly, like depression and Alzheimer's disease (which fall under the macro categories of mental disorders and nervous system diseases, respectively), but also with other relevant health conditions. For example, subjects affected by chronic bronchitis, osteoporosis, gastroenteritis, and many other pathologies have a higher median FI compared to those not affected. These conditions, although not directly included in the computation of the FI, still emerge as factors correlated with the subjects' level of frailty.

These results confirm that the FI, despite being constructed from a limited set of variables, manages to reflect a broader picture of the general health status of the population, demonstrating high sensitivity and predictive capacity.

Figure 5: Median, 1st and 3rd quartile of FI in subjects in the presence and absence of a disease or condition.

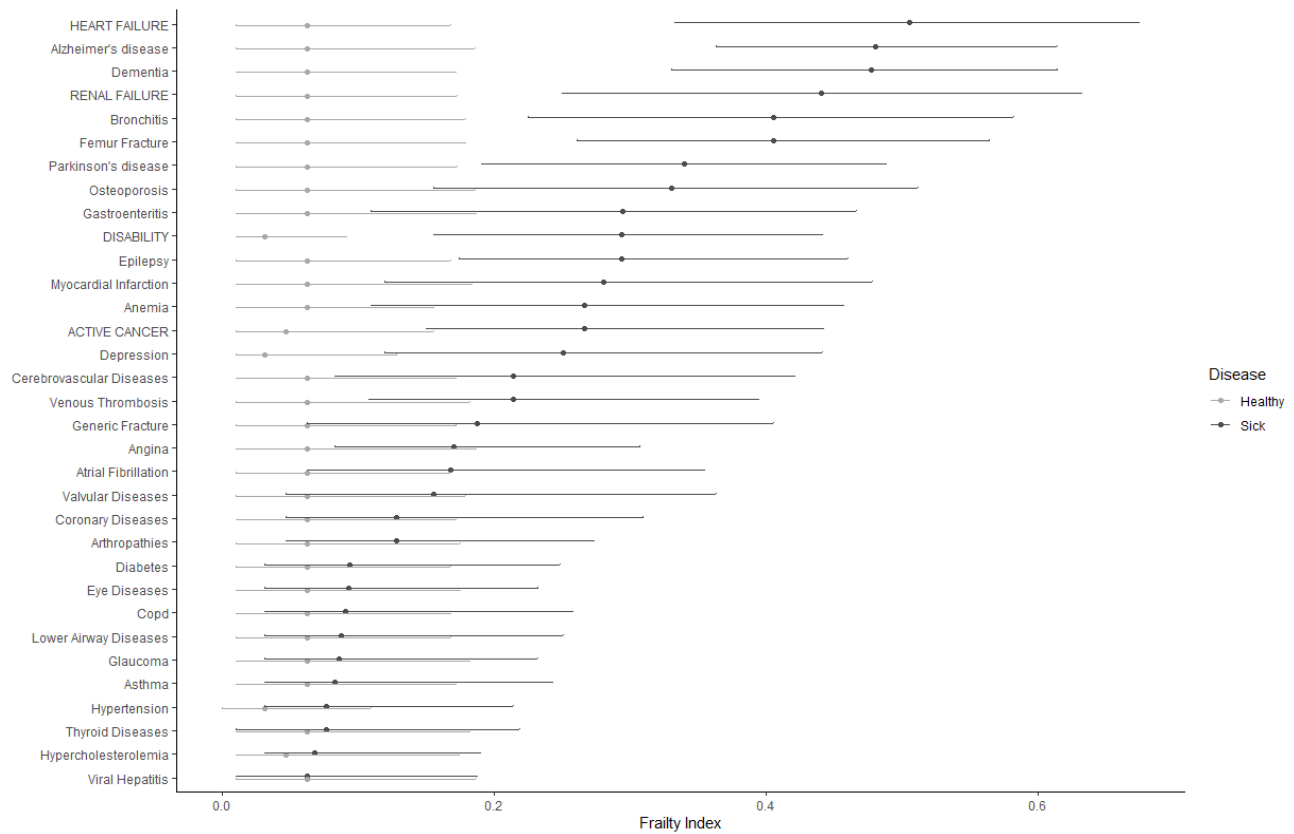

*Frailty Index and comorbidity*

Comorbidity is defined as the existence or occurrence of any additional disease, disorder, or pathological condition in the clinical course of a patient with a disease. The Charlson Comorbidity Index (CCI) is a widely used tool to assess the weight of comorbidities on an individual's probability of survival in the medium to long term [35]. The index considers a list of predefined pathologies (congestive heart failure, peripheral vascular disease, chronic pulmonary disease, liver disease, diabetes, myocardial infarction, cerebrovascular diseases, rheumatic and connective tissue diseases, peptic ulcer, hemiplegia, leukemia, lymphomas, kidney diseases, and tumours), each of which is associated with a score (from 1 to 6) reflecting the severity and impact of the condition on the patient's prognosis. The sum of the scores obtained for each pathology constitutes the overall value of the CCI.

For the purposes of the present study, the Index was calculated according to the adaptation of Deyo and co-authors so that it could be used with recorded ICD-9-CM diagnoses [36]. Having administrative data available, we calculated the Index for each hospitalisation over the course of a year, considering only the concomitant diagnoses reported in hospital discharge records. The CCI value assigned to the subject corresponds to the maximum recorded among all hospitalisations of the year.

The CCI has been collapsed into 4 classes: 0 (absence of comorbidities); 1 (maximum one mild comorbidity); 2 (maximum one moderate comorbidity or two mild comorbidities); and 3 or higher (representing severe or critical comorbidities or complex situations with several coexisting comorbidities).

Figure 6 shows the means and related 95% confidence intervals of the FI for subjects belonging to different classes of CCI. These results provide an overview of the association between comorbidity level and frailty. Figure 6 clearly indicates a direct relationship between the level of comorbidity and individual frailty. This trend confirms that subjects with a higher burden of diseases tend to be characterised by a higher level of frailty.

Figure 6: Means and 95% confidence intervals of FI for subjects belonging to different classes of CCI.

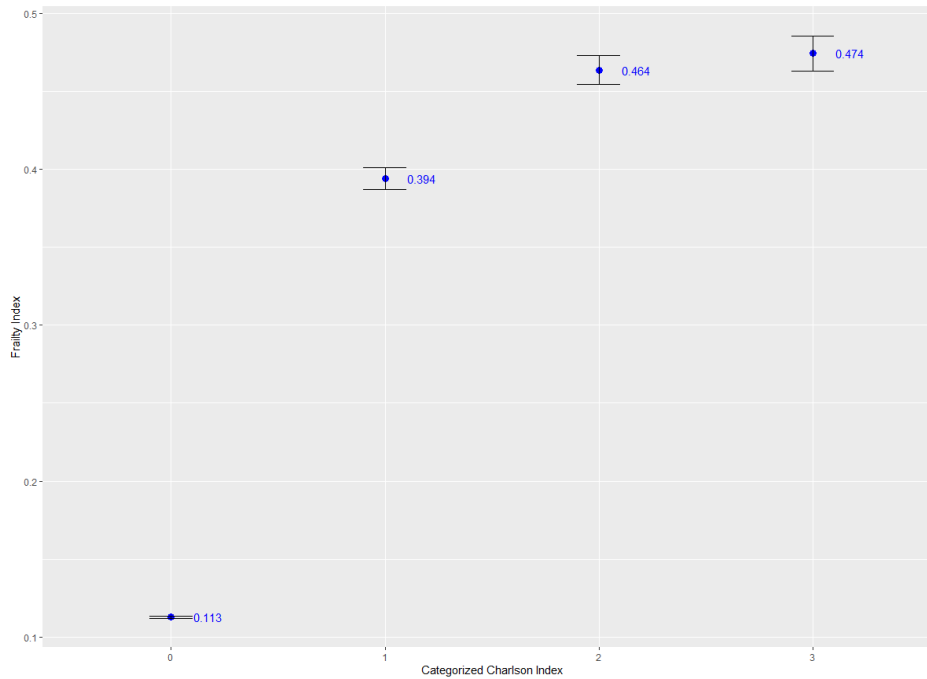

*Frailty, disability and comorbidity*

Frailty, comorbidity, and disability are distinct but interrelated concepts in geriatric health [37]. Comorbidity refers to the occurrence of additional pathological conditions that exacerbate the clinical course of a patient with a disease, while disability involves reduced ability to perform daily activities [38, 39]. Frailty is characterised by a decrease in physiological reserve and increased vulnerability to stressors. The relationship between these concepts is complex: frailty and comorbidity can be both predictors and outcomes of each other, while both contribute to disability [40]. Research suggests that frailty could be a better predictor of mortality than disability or comorbidity alone [41].

Figure 7 illustrates these interrelationships showing the percentage of the total population in a Venn diagram, based on data from the 2018 cohort. For this analysis, we define as frail the top 10% of the population according to our FI. Comorbidity was defined as a Charlson Index score equal to or higher than 1, a condition that affects 3.83% of the population. Disability was found in 23.47% of the total population.

In the figure, we can observe that 1.93% of the population simultaneously presents all three conditions, identifying a high-risk group that requires intensive and multidisciplinary clinical attention. 6.76% of the population shows an overlap between frailty and disability. This suggests that frailty and disability often coexist, potentially influencing each other. In contrast, only 0.33% of the cohort presents a combination of comorbidity and frailty without disability, indicating that when these two conditions coexist, they are often also associated with functional limitations. It is particularly interesting to note that only one-tenth of subjects classified as frail by FI do not present comorbidity or disability. These data underscore the complex nature of frailty and its close relationship with the other two conditions.

Figure 7: Venn diagram of frailty, disability, and comorbidity conditions in the 2018 cohort (prevalence in the 2018 cohort).

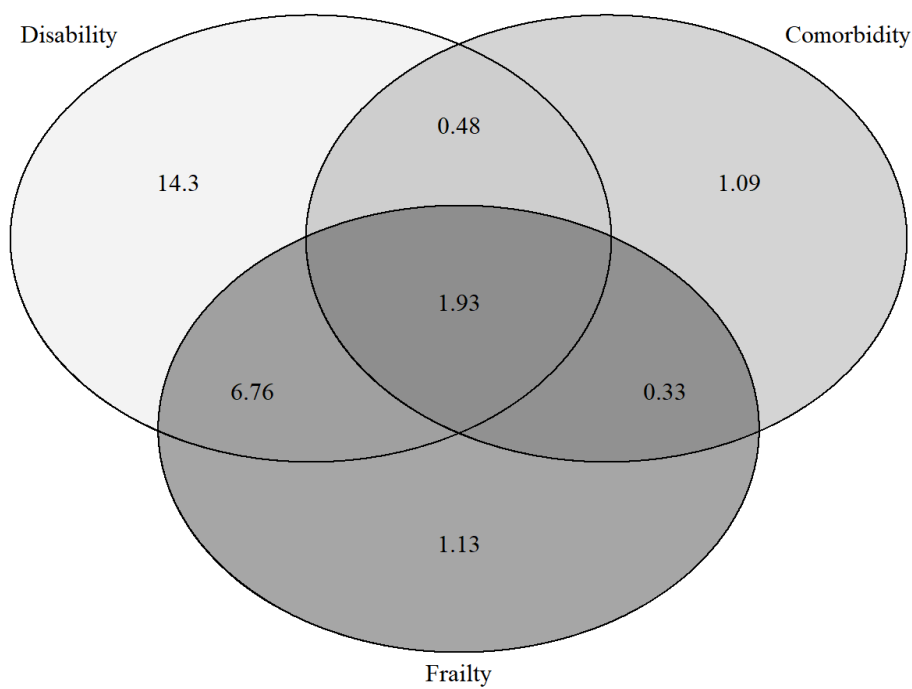

*Frailty and socioeconomic deprivation*

As the FI is based on administrative health data, individual socioeconomic characteristics are not available. To address this limitation, subjects were associated to their respective geographical areas of residence. This allowed us to assign to each subject the socioeconomic deprivation index proposed by Caranci and coauthors, which is a tool used to measure the level of socioeconomic disadvantage in each small geographical area [42]. The deprivation index (DI) is constructed considering five factors: the percentage of the population with education equal to or lower than elementary school; the percentage of the active population unemployed or seeking first employment; the percentage of occupied rented dwellings; the percentage of single-parent families; and the housing density (measured as occupants per $100m^2$).

The deprivation index is often divided into population quintiles. Figure 8 shows the mean and 95% confidence intervals of the FI for the 2018 cohort, divided by the quintiles of the deprivation index. The figure shows that as the level of deprivation increases, the FI also increases. Indeed, subjects belonging to the most disadvantaged quintiles (4th and 5th quintile) have, on average, higher FI compared to those in the less disadvantaged quintiles.

These results highlight a direct relationship between unfavourable socioeconomic conditions and greater vulnerability, suggesting that the social and environmental context plays a significant role in determining people's health status and need for assistance.

Figure 8: Means and 95% confidence intervals of FI for subjects belonging to different quintile groups of the deprivation index at the regional level.

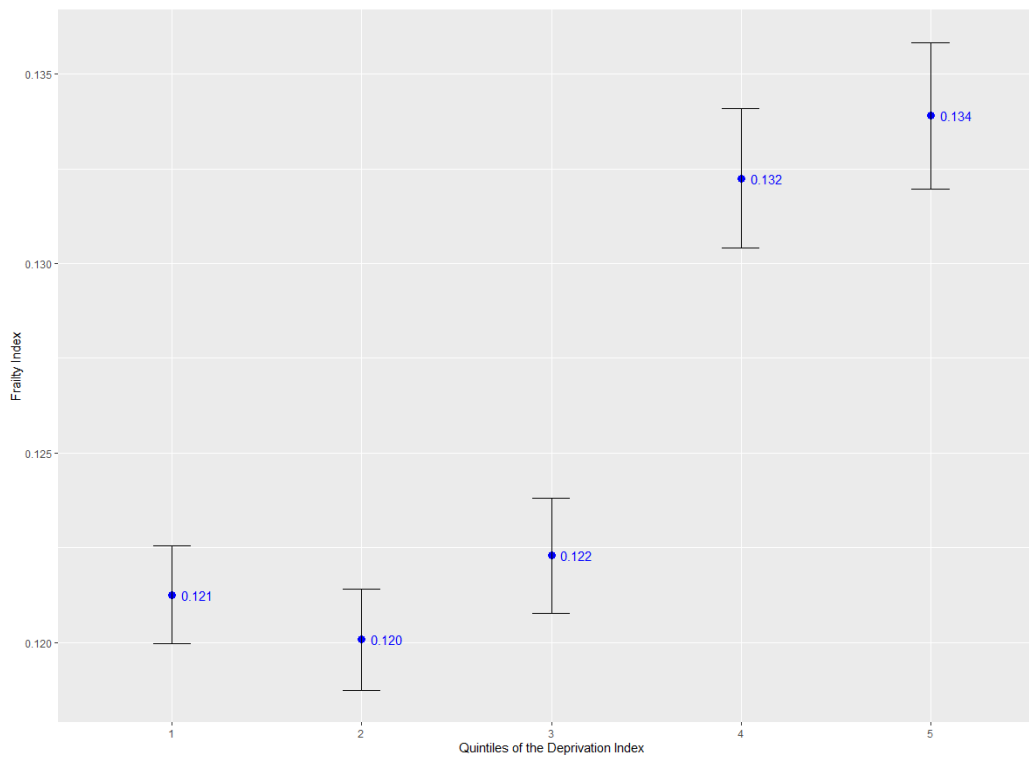

## DISCUSSION

### *Key Findings*

In this paper, we have proposed and validated a frailty index based on administrative health data, with the objective of stratifying the population according to the level of frailty in health. The FI is composed of a small number of variables, only eight (age in classes, presence of disability, number of hospitalisations, psychiatric disorders, nervous system diseases, heart failure, renal failure, cancer), but nevertheless it has very good performance, measured by the AUC, in predicting some frailty outcomes (death, access to the emergency room with highest priority, onset of dementia) and good performance for others (onset of disability and femur fracture). Its predictive ability for hospitalisation, which has a high prevalence and is a much more nonspecific outcome compared to the others, is not as good.

The results show a strong association between the highest levels of frailty and the frequency of the most severe outcomes, such as death, disability, and dementia. We observe, for example, that more than a third of those who are in the most severe 1% died the year after the calculation of the FI and more than half were hospitalised, or that within the most severe 5% of subjects, more than 90% are composed of disabled people. Even extending the analysis to a broader portion, such as 10% of the most frail, a progressive decrease in severe outcomes is observed, but maintaining higher levels compared to the rest of the population.

The analysis by quartiles further confirmed the validity of the index, showing that outcomes are predominantly concentrated in the highest quartiles of FI. In particular, while the last quartile gathers most of the severe events, in the intermediate quartiles there are subjects with more heterogeneous profiles, including both people with less severe conditions and subjects at moderate risk. Moreover, the first quartile is mainly populated by young individuals in good health conditions, with a very low or null number of outcomes.

Although the frailty indicator is based on only eight variables, it also discriminates well based on other chronic diseases.

Diseases that appear to be the mainly related to frailty are heart failure, Alzheimer's disease, dementia, renal failure, bronchitis, femur fracture, and Parkinson's disease. More generally, it is possible to create a ranking of diseases according to the level of frailty associated with them. Moreover, the analysis also shows a direct relation between frailty and comorbidity.

It is also interesting to note that higher levels of socioeconomic deprivation are associated with higher levels of health frailty.

*Strengths and limitations*

The strength of the approach proposed here lies in the final simplicity of calculation, thanks to the small number of variables considered, and its very satisfactory performance in terms of predicting ability, even in comparison to what is found in the literature, where indexes based on many more variables are proposed, but with similar or even worse performance.

The POSET approach is such that it is not necessary to calculate weights for each variable; the FI is regenerated each time. The only underlying assumption is that the variables that make up the FI remain the same. Several robustness analyses have been carried out on this aspect, which have confirmed the solidity of the approach.

From a computational point of view, the calculation time of the FI is absolutely acceptable.

Another strength is the use of administrative data only, available in every Local Health Unit. The data we used are drawn from administrative flows available in Italy and abroad.

Moving on to the weaknesses, we can say that the limitation of POSET is that it is population dependent. If the fact of regenerating the FI each time is a strength of the method, this can at the same time constitute a weakness, in the sense that, as the population changes, the same subject in the same conditions may have slightly different values of the FI. From a practical point of view, especially when working on large populations that cover almost all possible profiles, this is not a problem: for the same profiles, we have verified that from one year to the next, the percentage of those who change the value of the FI (based on sensitivity intervals calculated with resampling methods) is 0.03%. From a methodological point of view, the issue remains, so we are studying an approach to arrive at a methodological proposal for calculating confidence intervals for the average rank calculated with the POSET approach.

*Use of administrative health data*

The points of attention on the use of administrative health data are largely related to the sharing of algorithms for the calculation of the variables used in the process, that is, chronic diseases, health conditions, and adverse events. For each of these variables, an algorithm has been developed (we have made available in the Supplementary Materials the algorithms of the variables that make up the FI; the other algorithms are available upon request from the authors), which considers the contents of each health flow and the coding systems adopted in each one, and extracts the cases for which the relevant coding for the disease under analysis have been selected in at least one of the flows considered. Some flows use internationally shared coding systems (e.g., diseases in hospital discharge records are coded with ICD-9-CM), others at the national level. The process of generating algorithms for each disease/condition is complex and is the result of joint work between physicians and statisticians. In Italy, the Italian Association of Epidemiology has made considerable effort in producing algorithms for various diseases, but not yet for all of them [43, 44]. Therefore, where possible, we have relied on the literature. In the absence of proposals, we have created algorithms ex novo with the support of experts. This aspect deserves, in our opinion, great attention, because the use of different definitions and algorithms necessarily leads to different and non-comparable results. This aspect has partly conditioned the choices we made, for example, on the frailty outcomes considered: in the results we show how hospitalisation is a rather nonspecific outcome, and therefore more targeted outcomes, such as urgent or repeated hospitalisation, might have better suited our case. Unfortunately, in Italy, such outcomes, not explicitly defined in classification systems, are interpreted differently in regional health systems, and therefore we decided not to use them.

The choice of which data sources to use is also crucial and must be shared. In this work, we have considered a wide range of health flows from which to draw information, and subsequent validations have shown that some of these flows have no particular impact (e.g., the psychiatric flow), while others are indispensable and have a different impact. For example, it changes the aggregate of subjects with diseases depending on whether only hospital discharges or drug use are considered. These two sources are crucial for identifying subjects with diseases, and their contribution is clearly different: the source on hospital discharges identifies subjects who are obviously more severe, while the pharmaceutical source includes subjects with a disease, not necessarily severe and managed pharmacologically. For example, we have verified the value of FI of subjects identified as diabetic using only admissions and emergency rooms, and using all flows that trace diabetes, including pharmaceuticals and exemptions. In the first case, the median value of the FI is 0.443, in the second case 0.178, showing that a different definition of the condition may strongly affect the analysis.

It is evident how crucial the choice of data sources and algorithms for constructing variables is in the final results. On this aspect, there needs to be strong awareness and sharing of tools and methods.

Although administrative health data are a valuable source, they have the limitation of not containing socioeconomic information about the subjects, nor measures of functional capacity. In the case of frailty, this is clearly a limitation, as we cannot directly measure frailty with the biomedical paradigm of Fried et al. or the bio-psycho-social paradigm of Gobbens et al., which can only be applied through direct interviews with the subjects [1, 15]. In fact, all proposals based on administrative health data follow the path of constructing measures that best predict adverse health outcomes. The tool we provide is therefore a measure of health frailty. Using the deprivation index at the census tract level, we have shown that there is a strong relationship between health frailty and socioeconomic deprivation. The study of such relationships and the causal links with social, economic, and environmental aspects cannot be addressed with administrative population data (at least in Italy, where there are limitations related to privacy regulations), but with data from specific surveys. An example is the Survey of Health, Ageing and Retirement in Europe (SHARE) and similar surveys conducted in many countries [45].

*Implications for research, policy, and practice*

The genesis of this study can be traced back to direct requests from epidemiologists in local health authorities. The request was made after evidence of the impact, mainly on frail older people, of the increasingly numerous summer heat waves. To this was added the COVID epidemic, which had a much more direct impact on frail older adults and made it clear that it was necessary to have tools that would allow the health decision-maker to act quickly and fully, being able to fast identify frail subjects and activate targeted actions.

The legislature has also reached these conclusions several times, indicating the need to make population stratification procedures available, as already illustrated in the Introduction of the article and as strongly hoped for in the National Recovery and Resilience Plan (PNRR).

From an operational point of view, the result of our research can become a simple-to-use tool for those working in the epidemiology or management control offices of Local Health Authorities.

However, we highlight the limitations that currently Italian legislation imposes, which make it very difficult to actually use such tools. In fact, stratification procedures are carried out on anonymised data, from which it is not possible to trace back to the individual. Currently, the results can only be analysed on an aggregate level.

**CONCLUSIONS AND FUTURE WORK**

In this paper, we have presented a pathway for the construction of a Frailty Index based on administrative health data. This is an increasingly relevant issue for public health management from a prevention and appropriate healthcare perspective.

To effectively use our proposal to the health decision-maker, we have prepared a tool based on open statistical software, through which it will be possible, once the original data are loaded, to produce the FI for all subjects and a series of tabular and graphical summaries. The tool is currently being validated on real data from an Italian province and region.

From a methodological point of view, we are proceeding with the proposal of confidence intervals based on POSET. This step forward will allow the comparison of the FI of the same subjects over time, or the comparison of different populations. The research question that needs to be answered in this case is to be able to disentangle to what extent significant variations in the level of frailty of the subjects are induced by actual changes in the state of frailty and not resulting from mere modifications of the structure of the set of profiles of the population.

## SUPPLEMENTARY MATERIALS

Table S.1: Operationalisation of the variables that make up the frailty index according to administrative health data, codes, and description.

| Variabile | Administrative health data | Codes |
|---|---|---|
| Age (65-69, 70-74, 75-79, 80-84, 85-89, 90+) | Regional health registry | |
| Total number of hospitalisations (0, 1-2, 3 or more times) | Hospital discharge records | From hospital discharge records, the number of hospitalisations of all types is counted. |
| Mental disorders (dichotomous, 1 if present) | Ticket exemptions | 005, 014, 044 |
| | Hospital discharge records, emergency room admissions | 291 – 319 (excluding 294.1 as second diagnosis) |
| | Territorial psychiatry | F04, F06 – F99 |
| | Territorial pharmaceuticals | N05, N06A, N07B |
| Neurological diseases (dichotomous, 1 if present) | Ticket exemptions | 011, 017, 029, 034, 038 with diagnosis: 332, 041, 046 |
| | Hospital discharge records, emergency room admissions | 320 – 359, 290, 294.1 (as a second diagnosis) |
| | Territorial psychiatry | F00 – F03, F05 |
| | Territorial pharmaceuticals | N03, N04, N06D |
| Cancer (dichotomous, 1 if present) | Ticket exemptions | 048 |
| | Hospital discharge records, emergency room admissions | 140 – 208, V10, V580, V581, V671, V672 |
| | Territorial pharmaceuticals | L01 |
| Disability (dichotomous, 1 if present) | Ticket exemptions | 3C1, 3C2, 3G1, 3G2, 3L1, 3L2, 3L3, 3M1, 3M2, 3M3, INAIL, G01, G02, L01, L02, L03, L04, S01, S02, S03, C01, C02, C03, C04 |

|  | Integrated home care | It looks at whether the individual has received at least one integrated home care intervention |
| --- | --- | --- |
| Heart failure (dichotomous, 1 if present) | Ticket exemptions | 021 |
|  | Hospital discharge records, emergency room admissions | 428 |
| Kidney failure (dichotomous, 1 if present) | Ticket exemptions | 023 |
|  | Hospital discharge records, emergency room admissions | 584 – 586 |

**AUTHORS CONTRIBUTION**

Giovanna Boccuzzo obtained funding for the study.
All Authors conceptualised and designed the analyses, and carried out the literature search.
Maurizio Nicolaio conducted the analyses, generated tables and figures.
Giovanna Boccuzzo wrote Introduction, Discussion and Conclusion Sections.
Maurizio Nicolaio and Margherita Silan wrote Results Section.
Margherita Silan wrote Methods Section
All Authors contributed to the development of the methodology, and interpretation of analyses and reviewed and revised the submitted manuscript.
All authors had final approval of the submitted manuscript.
The corresponding author attests that all the authors listed meet authorship criteria and that no others who meet the criteria have been omitted.

**ACKNOWLEDGMENTS**

The authors acknowledge Next Generation EU, in the context of the National Recovery and Resilience Plan, Investment PE8 – Project Age-It: "Ageing Well in an Ageing Society", CUP C93C22005240007 [DM 1557 11.10.2022]. The views and opinions expressed are only those of the authors and do not necessarily reflect those of the European Union or the European Commission. Neither the European Union nor the European Commission can be held responsible for them.
The authors would also thank ULSS6 Euganea (Padua province, Italy), in particular the Prevention office, with which the Department of Statistical Sciences of the University of Padua activated the StHeP (State of Health in Padua) convention, for the work of data extraction and anonymisation.

**STATEMENT OF CONFLICT OF INTERESTS**

None declared.

**ETHICS STATEMENT**

As only routinely acquired de-identified data were analysed no research ethics committee approval was required by the Health Research Authority. Anonymised data were made available by the Veneto Region Local Health Unit 6 (LHU6) of the Veneto region (Italy) within the StHeP agreement (State of Health in Padua) between LHU6 and the Department of Statistical Sciences of the University of Padua. The data was subjected to a rigorous anonymization process, with no chance of individuals being identifiable, as required by Italian legislation for the use of data for scientific research purposes.

**DATA AVAILABILITY STATEMENT**

The data supporting this study's findings are held by LHU6 and were used under license for this work, but they are not available to the general public.

**ABBREVIATIONS**

AR: Average Rank
AUC: Area Under the Curve
CCI: Charlson Comorbidity Index
ER: Emergency Room
EU: European Union
FI: Frailty Index
LHU: Local Health Unit
PNRR: National Recovery and Resilience Plan
POSET: Partially Ordered Set
ROC: Receiver Operating Characteristic curve
TFI: Tilburg Frailty Indicator